\documentclass[%
reprint,
final,
superscriptaddress,
nofootinbib,
nobibnotes,
amsmath,amssymb,
aps,
prl,
]{revtex4-2}

\usepackage{graphicx}% Include figure files
\usepackage{dcolumn}% Align table columns on decimal point
\usepackage{bm}% bold math
\usepackage[normalem]{ulem}  % used to strikeout text 

\usepackage{dsfont}
\usepackage{color,colortbl,xcolor}
\usepackage{layouts}
\newcommand{\Nabla}{\vec{\nabla}}%
\newcommand{\norm}[1]{\lVert#1\rVert}%
\newcommand{\Peclet}{P\'eclet}%
\newcommand{\tlj}{\tau_\mathrm{LJ}}
\allowdisplaybreaks
\newcommand{\sigrep}{a}% Command to replace diameter of particles sigma

\renewcommand{\vec}[1]{\bm{#1}}
\newcommand{\hatvec}[1]{\hat{\vec{#1}}}

\begin{document}

\title{Orientation-dependent propulsion of active Brownian spheres: from self-advection to programmable cluster shapes}

\author{Stephan Br\"oker}
\affiliation{Institut f\"ur Theoretische Physik, Center for Soft Nanoscience, Westf\"alische Wilhelms-Universit\"at M\"unster, 48149 M\"unster, Germany}

\author{Jens Bickmann}
\affiliation{Institut f\"ur Theoretische Physik, Center for Soft Nanoscience, Westf\"alische Wilhelms-Universit\"at M\"unster, 48149 M\"unster, Germany}

\author{Michael te Vrugt}
\affiliation{Institut f\"ur Theoretische Physik, Center for Soft Nanoscience, Westf\"alische Wilhelms-Universit\"at M\"unster, 48149 M\"unster, Germany}

\author{Michael E. Cates}
\affiliation{DAMTP, Centre for Mathematical Sciences, University of Cambridge, Cambridge CB3 0WA, United Kingdom}

\author{Raphael Wittkowski}
\email[Corresponding author: ]{raphael.wittkowski@uni-muenster.de}
\affiliation{Institut f\"ur Theoretische Physik, Center for Soft Nanoscience, Westf\"alische Wilhelms-Universit\"at M\"unster, 48149 M\"unster, Germany}

\begin{abstract}
Applications of active particles require a method for controlling their dynamics. While this is typically achieved via direct interventions, indirect interventions based, e.g., on an orientation-dependent self-propulsion speed of the particles, become increasingly popular. In this work, we investigate systems of interacting active Brownian spheres in two spatial dimensions with orientation-dependent propulsion using analytical modeling and Brownian dynamics simulations. It is found that the orientation-dependence leads to self-advection, circulating currents, and programmable cluster shapes.
\end{abstract}
\maketitle

Active Brownian particles (ABPs) \cite{Elgeti2015,Speck2016,Zttl2016,BechingerEA16} combine Brownian motion with directed self-propulsion, leading to an inherently nonequilibrium dynamics. They are a prime model system for active particles, which have great potential for future applications including nanobots for medical applications like microsurgery \cite{Kurinomaru2020} or drug delivery \cite{Alapan2018,Yasa2018,Ceylan2019} and programmable materials for industrial applications \cite{Tibbits2014,Jones2015,Qi2020}. Almost all applications have in common that general features of the dynamics of active particles, such as their collective dynamics, have to be controlled. This is often achieved using direct interventions \cite{BechingerEA16,BickmannBroekerW2022external}, where an external force or torque acts on the particles. Recently, methods based on indirect interventions, where one instead changes the way the particles perceive their environment, have become very popular. Previous work on such approaches focuses on motility maps, where the particles' propulsion speed becomes space-dependent \cite{Magiera2015,Lozano2016,Stenhammar2016,Geiseler2017,Sharma2017,Grauer2018,Liebchen2019,Lozano2019propagating,Lozano2019diffusing,CapriniMWL2022,CapriniMWL2022b,ArltMDPP2018,FrangipaneDPMSBVBD2018,ArltMDPP2019}. Such systems have already been realized, e.g., via light-propelled particles in complex light fields \cite{Dai2016,Lozano2016}. Less well understood are indirect interventions with respect to the particles' orientations, as given, e.g., by an orientation-dependent propulsion force. Such forces arise, e.g., when particles are propelled by ultrasound \cite{VossW2022} or light \cite{JeggleRDW2022}.

There exists theoretical as well as experimental work on single particles with an orientation-dependent self-propulsion \cite{Uspal2019,sprenger2020active}, but many-particle systems of interacting ABPs with an orientation-dependent propulsion have not been investigated so far. Of particular importance in this context are the effects of  such an indirect intervention on the collective dynamics of ABPs and their intriguing nonequilibrium effects, such as non-state-function pressure \cite{Solon2015Nat,Solon2015a}, reversed Ostwald ripening \cite{Tjhung2018}, and motility-induced phase separation (MIPS) \cite{MIPS}.

In this article, we address this issue by investigating systems of interacting spherical ABPs with an orientation-dependent propulsion velocity in two spatial dimensions using analytical modeling and computer simulations. We derive a predictive field-theoretical model that describes the collective dynamics of such systems and find novel contributions that depend on the symmetry properties of the orientation-dependent propulsion. The model provides an analytical prediction for the spinodal corresponding to the onset of MIPS, which we compare to state diagrams obtained by Brownian dynamics simulations. Furthermore, we show that the orientation dependence of the propulsion gives rise to the self-assembly of deformed MIPS clusters with, e.g., elliptical, triangular, and rectangular shapes.

The considered system consists of $N$ spherical, interacting ABPs in two spatial dimensions with center-of-mass positions $\vec{r}_i = (x_{i}, y_{i})^\mathrm{T}$, orientations $\hatvec{u}(\phi_i) = (\cos(\phi_i), \sin(\phi_i))^\mathrm{T}$, and polar orientation angles $\phi_i$, where $i = 1,\dots,N$.  To model the microscopic dynamics of the particles, we use the overdamped Langevin equations
\begin{align}
\dot{\vec{r}}_i &= v_\mathrm{A}(\phi_i)\hatvec{u}(\phi_i)+\vec{v}_{\mathrm{int}, i}(\lbrace \vec{r}_i\rbrace)+\sqrt{2D_\mathrm{T}}\vec{\Lambda}_{\mathrm{T},i},\label{eqn:LangevinR}\\
\dot{\phi}_i &= \sqrt{2D_\mathrm{R}}\Lambda_{\mathrm{R},i},\label{eqn:LangevinPHI}
\end{align}
where an overdot denotes a derivative with respect to time $t$. Equations \eqref{eqn:LangevinR} and \eqref{eqn:LangevinPHI} differ from the standard Langevin equations for ABPs \cite{Solon2015a,Speck2016,Zttl2016,BickmannW2020,BickmannBJW2020,BickmannBroekerW2022external} by the orientation-dependence of the propulsion speed $v_\mathrm{A}(\phi)$. Particle interactions are incorporated using the term $\vec{v}_{\mathrm{int}, i}(\lbrace\vec{r}_i\rbrace) = -\beta D_\mathrm{T}\sum_{j=1, j\neq i}^N \Nabla_{\vec{r}_i} U_2(\norm{ \vec{r}_i-\vec{r}_j})$. Here, $\beta = 1/(k_\mathrm{B} T)$ is the thermodynamic beta with Boltzmann constant $k_\mathrm{B}$ and temperature $T$, $D_\mathrm{T}$ the translational diffusion coefficient, $\Nabla_{\vec{r}_i} = (\partial_{x_{i}}, \partial_{y_{i}})^\mathrm{T}$ the del operator with respect to $\vec{r}_i$, $U_2$ a two-particle interaction potential, $\norm{\cdot}$ the Euclidean norm, $D_\mathrm{R}=3D_{\mathrm{T}}/a^2$ the rotational diffusion coefficient, and $a$ the particle diameter. Thermal fluctuations are modeled via zero-mean, unit-variance statistical white noises $\vec{\Lambda}_{\mathrm{T}, i}(t)$ and $\Lambda_{\mathrm{R},i}(t)$.
 
Using the interaction-expansion method \cite{RW,BickmannBJW2020,BickmannW2020,BickmannW2020b,teVrugtFHHTW2022,teVrugtBW2022}, we derived from Eqs.\ \eqref{eqn:LangevinR} and \eqref{eqn:LangevinPHI} an advection-diffusion model that describes the time evolution of the number density $\rho(\vec{r}, t)$ of the particles, depending on position $\vec{r}=(x,y)^\mathrm{T}$ and time. The derivation (see Ref.\ \cite{SupMat}) assumes short-ranged interactions and a dependence of $v_\mathrm{A}(\phi)$ on $\phi$ that can be well approximated with few Fourier modes. The resulting model reads
\begin{equation}
\dot{\rho} = - \Nabla \cdot \big(\vec{\mu}^{(1)} \rho\big) + \Nabla \cdot\big(\underline{D}(\rho) \Nabla \rho\big),\label{eqn:Continuity_equation}
\end{equation}
where $\underline{D}(\rho)$ is a density-dependent diffusion tensor with elements
\begin{equation}
D_{ij}(\rho) = (D_\mathrm{T} + c_1 \rho + c_2 \rho^2)\delta_{ij} + c_3 \rho \mu^{(2)}_{ij} +  \frac{\mu^{(2)}_{ik}\mu^{(2)}_{kj}}{2 D_{\mathrm{R}}}. \label{eqn:D_tensor}
\end{equation}
(We sum over repeated lower indices from here on.) The coefficients $c_i$ are given in Ref.\ \cite{SupMat}. Moreover, $\delta_{ij}$ denotes the Kronecker delta, $\vec{\mu}^{(1)} =\int_0^{2\pi}\!\!\mathrm{d}\phi \ v_\mathrm{A}(\phi) {\hatvec{u}}(\phi)/(2\pi)$ the particles' orientation-averaged propulsion velocity, and $\mu^{(2)}_{ij}$ the elements of the symmetric velocity tensor $\underline{\mu}^{(2)} = \int_0^{2\pi}\!\!\mathrm{d}\phi \, v_\mathrm{A}(\phi) {\hatvec{u}}(\phi)\otimes{\hatvec{u}}(\phi)/\pi$ with dyadic product $\otimes$. Formally, $\vec{\mu}^{(1)}$ and $\underline{\mu}^{(2)}$ are the zeroth- and first-order contributions, respectively, of the orientational expansion of the propulsion velocity $\vec{v}(\phi)=v_{\mathrm{A}}(\phi)\hatvec{u}(\phi)$ into Cartesian tensors \cite{TeVrugtW19,teVrugtW2019c}: $\vec{v}(\phi) = \vec{\mu}^{(1)} + \hatvec{u}(\phi)\cdot\underline{\mu}^{(2)} + \mathcal{O}(\hatvec{u}^2)$. Including the $\mathcal{O}(\hatvec{u}^2)$ contributions would not lead to additional terms in Eq.\ \eqref{eqn:Continuity_equation} unless we also include higher orders in derivatives. For the special case of an isotropic propulsion speed, we recover the purely diffusive model from Ref.\ \cite{BickmannW2020}. Equation \eqref{eqn:Continuity_equation} is our first main result. 
 
The orientational contributions $\vec{\mu}^{(1)}$ and $\underline{\mu}^{(2)}$ are the novel features of this model compared to the model derived in Ref.\ \cite{BickmannW2020} for isotropic propulsion. A nonvanishing $\vec{\mu}^{(1)}$ corresponds to an internal polarization of the propulsion velocity that, similar to an external field \cite{BickmannBroekerW2022external}, gives rise to (self-)advection. This is easily seen from the fact that the first term on the right-hand side of Eq.\ \eqref{eqn:Continuity_equation} can be eliminated using the Galilei transformation $\vec{r} \to \vec{r} - \vec{\mu}^{(1)}t$. The self-advection results, like the motility of individual active particles \cite{Walther2008,Walther2013,voss2020shape}, from an $(\vec{r}\leftrightarrow -\vec{r})$-symmetry breaking. In contrast, $\underline{\mu}^{(2)}$ breaks the $(x \leftrightarrow y)$-symmetry of the diffusion tensor \eqref{eqn:D_tensor}. This also occurs in systems with chirality such as circle swimmer systems \cite{liao2018clustering,BickmannBJW2020}.
 
For what follows, we specify the orientation-dependent propulsion speed $v_{\mathrm{A}}(\phi)$ as
\begin{equation}
v_{\mathrm{A}, n}(\phi) = \bar{v}\big(1 - \nu + 2 \nu \sin^2(n \phi/2)\big), 
\label{eqn:v_A_n_fold}%
\end{equation}
which involves an $n$-fold symmetry and is parametrized by the orientation-averaged propulsion speed $\bar{v}$ and the dimensionless angular modulation amplitude $\nu$. A similar form was used in Ref.\ \cite{sprenger2020active}.
We focus on the cases $n = 1,\dots,4$. 
Using Eq.\ \eqref{eqn:v_A_n_fold}, we obtain
\begin{align}
\vec{\mu}^{(1)} &= -\bar{v}\delta_{n, 1} \frac{1}{2}( \nu, 0)^\mathrm{T}, \label{eqn:Explicit1FS} \\
\underline{\mu}^{(2)} &= \bar{v}\Big(\mathds{1}   - \underline{\sigma}_{3} \delta_{n, 2} \frac{\nu}{2}\Big), \label{eqn:Explicit2FS}
\end{align}
where $\mathds{1}$ is the identity matrix and 
$\underline{\sigma}_{3}$ is the third Pauli matrix.
Only for $n=1$, $\vec{\mu}^{(1)}$ does not vanish. The tensor $\underline{\mu}^{(2)}$ is diagonal and nonzero for all $n$. For $n=2$, its diagonal elements are not identical.

To investigate the system further, we performed Brownian dynamics simulations \cite{Plimpton1995} based on the Langevin equations \eqref{eqn:LangevinR} and \eqref{eqn:LangevinPHI}. For the interactions, we chose the Weeks-Chandler-Andersen potential $U_2(r)=(4\varepsilon [ ( \sigrep/r )^{12} - (\sigrep/r )^{6} ] + \varepsilon)\Theta(2^{1/6}a-r)$ \cite{WeeksCA1971} with interaction strength $\varepsilon$, particle distance $r$, and Heaviside step function $\Theta$. The particle diameter $a$, Lennard-Jones time scale $\tlj=a^2/(\varepsilon \beta D_\mathrm{T})$, and interaction strength $\varepsilon$ are chosen as units of length, time, and energy, respectively. 
Nondimensionalization of Eqs.\ \eqref{eqn:LangevinR} and \eqref{eqn:LangevinPHI} leads to the \Peclet{} number $\mathrm{Pe} = \bar{v} a/D_\mathrm{T}$, which is a measure for the activity of the particles, and the overall packing density $\Phi_0 = \pi\bar{\rho} \sigrep^2/4$, where $\bar{\rho}$ is the spatially-averaged number density of the particles. We fixed the average propulsion speed to $\bar{v} = 24 a / \tlj$ and changed $\mathrm{Pe}$ via the temperature $T$. If not stated otherwise, we chose $\mathrm{Pe} = 150$ and $\Phi_0 = 0.4$. Additional details on the computer simulations can be found in Ref.\ \cite{SupMat}.

\begin{figure}[htpb]
\centering
\includegraphics[width=\linewidth]{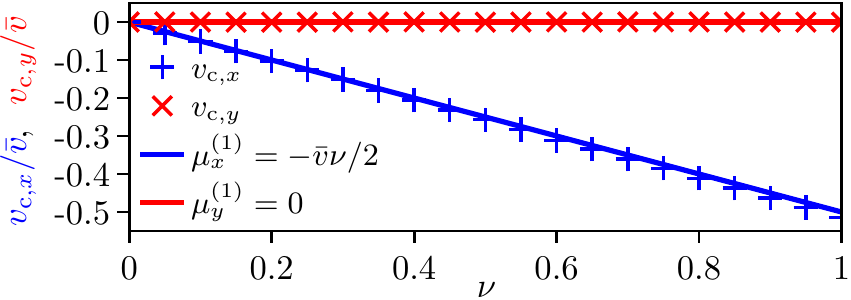}
\caption{\label{fig:1}Comparison of our analytical results for the self-advection velocity $\vec{\mu}^{(1)}=(\mu^{(1)}_x,\mu^{(1)}_y)^{\mathrm{T}}$ (see Eq.\ \eqref{eqn:Explicit1FS}) with the time-averaged velocity of particle clusters $\vec{v}_\mathrm{c} = (v_{\mathrm{c},x}, v_{\mathrm{c},y})^{\mathrm{T}}$ that we obtained from six Brownian dynamics simulations for each considered value of the angular modulation amplitude $\nu$.}
\end{figure}

Equation \eqref{eqn:Explicit1FS} predicts the self-advection velocity $\vec{\mu}^{(1)}$ of MIPS clusters for $n=1$. We confirmed this by comparing Eq.\ \eqref{eqn:Explicit1FS} with the velocity $\vec{v}_c = (v_{c,x}, v_{c,y})^\mathrm{T}$ of macroscopic MIPS clusters of ABPs that we observed in Brownian dynamics simulations.
Figure \ref{fig:1} shows that our analytical prediction and the velocity of the clusters are in excellent agreement, demonstrating that our theory is applicable also to self-organized structures. The $x$-component of $\vec{v}_c$ decreases linearly for increasing $\nu$ and the $y$-component is zero since $v_{\mathrm{A}, 1}(\phi) = v_{\mathrm{A}, 1}(-\phi)$.
Our results suggest that, by changing the first Fourier mode of $v_\mathrm{A}(\phi)$, one can steer many-particle structures of ABPs in arbitrary directions. This effect can be useful for applications where one wants, e.g., to steer clusters of ABPs through a maze or into or out of a trap, including active carrier particles that need to swim to certain regions and release their cargo there \cite{Alapan2018,Yasa2018,Ceylan2019}. The ability to steer ABPs collectively is relevant, e.g., for drug delivery \cite{NitschkeSW2021} and active microstructures \cite{Stenhammar2016}. 

\begin{figure*}[htpb]
\centering
\includegraphics[width=\linewidth]{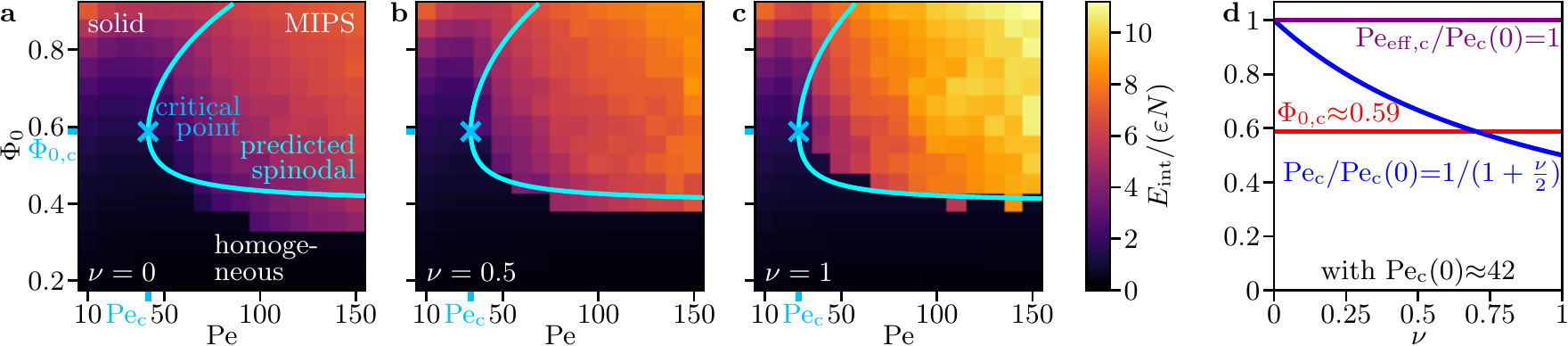}
\caption{\label{fig:2}(\textbf{a})-(\textbf{c}) State diagrams for ABPs with orientation-dependent propulsion speed $v_{\mathrm{A}, 2}(\phi)$ (see Eq.\ \eqref{eqn:v_A_n_fold}) showing the reduced average interaction energy per particle $E_{\mathrm{int}}/(\varepsilon N)$ as a function of the particles' \Peclet{} number $\mathrm{Pe}$ and overall packing density $\Phi_0$ for different values of the angular modulation amplitude $\nu$. The state diagrams include a homogeneous state, MIPS, and a solid state. Our analytical prediction for the spinodal corresponding to the onset of MIPS and the associated critical point are indicated by cyan curves and blue cross marks, respectively. (\textbf{d}) Predicted values of $\mathrm{Pe}$, $\mathrm{Pe}_{\mathrm{eff}}$, and $\Phi_0$ at the critical point (denoted $\mathrm{Pe}_{\mathrm{c}}$, $\mathrm{Pe}_{\mathrm{eff,c}}$, and $\Phi_{0,\mathrm{c}}$) as a function of $\nu$.}
\end{figure*}

For $n=2$, no self-advection arises, but the diffusion tensor has no longer identical diagonal elements: $D_{ij} \propto f(\rho)\delta_{ij} -\bar{v}\nu( c_3 \rho + \bar{v}/D_\mathrm{R} ) \sigma_{3, ij}$ with a scalar function $f(\rho)$. Since the diffusion tensor determines the spinodal for the onset of MIPS \cite{BickmannW2020}, the spinodal differs from the standard case. A linear stability analysis yields the spinodal condition $D_{22}(\rho) = 0$. This condition becomes equal to the one presented in Ref.\ \cite{BickmannW2020} for standard ABPs with isotropic propulsion when one replaces the ordinary \Peclet{} number there by a new effective \Peclet{} number $\mathrm{Pe}_{\mathrm{eff}} = \sigrep\, \textrm{max}(\mu^{(2)}_{11}, \mu^{(2)}_{22})/D_\mathrm{T} = (1 + \nu/2) \mathrm{Pe}$. To confirm the spinodal condition and this mapping, we obtained, for different angular modulation amplitudes $\nu$, state diagrams by Brownian dynamics simulations. We chose the reduced average interaction energy per particle $E_\mathrm{int}(\mathrm{Pe}, \Phi_0)/(\varepsilon N)$ \cite{BickmannBJW2020} as a quantity for the identification of clusters. The analytical prediction and simulation results for the spinodal are shown in Fig.\ \ref{fig:2}. 
 
For $\nu=0$, which corresponds to standard MIPS, the state diagram (see Fig.\ \ref{fig:2}a) shows a homogeneous state for low $\Phi_0$, a solid state for small $\mathrm{Pe}$ and large $\Phi_0$, and MIPS clusters for large $\mathrm{Pe}$ and moderate or large $\Phi_0$. Both of the latter two states have a higher reduced interaction energy per particle $E_\mathrm{int}/(\varepsilon N)$ than the homogeneous state. For increasing $\nu$, the state diagram gradually changes and the MIPS clusters emerge at lower $\mathrm{Pe}$ (see Figs.\ \ref{fig:2}b,c). The analytical prediction exhibits the same qualitative behavior and is in very good agreement with the simulation data even for large modulation amplitudes $\nu =1$. As shown in Fig.\ \ref{fig:2}d, while the value $\mathrm{Pe}_{\mathrm{c}}$ of $\mathrm{Pe}$ at the critical point decreases with $\nu$, the critical value $\mathrm{Pe}_{\mathrm{eff,c}}$ of $\mathrm{Pe}_{\mathrm{eff}}$ remains (like the critical value $\Phi_{0,\mathrm{c}}$ of $\Phi_0$) unchanged. This confirms our claim that one can map the phase-separation behavior onto that of standard ABPs by making use of an effective \Peclet{} number. The spinodal for MIPS is our second main result.
 
\begin{figure}[htpb]
\centering
\includegraphics[width=\linewidth]{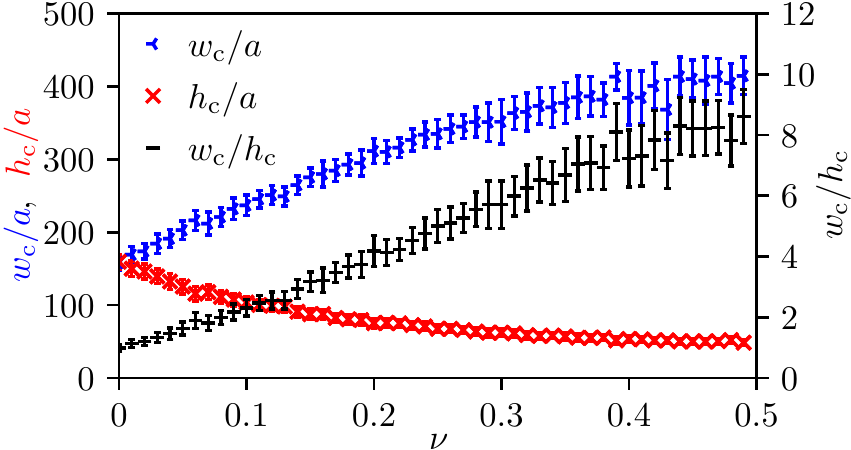}
\caption{\label{fig:3}Width $w_\mathrm{c}$, height $h_\mathrm{c}$, and aspect ratio $w_\mathrm{c}/h_\mathrm{c}$ of MIPS clusters consisting of ABPs with propulsion speed $v_{\mathrm{A}, 2}(\phi)$ (see Eq.\ \eqref{eqn:v_A_n_fold}) as functions of the angular modulation amplitude $\nu$.}
\end{figure}

For $n=2$, but not for $n=1$, we also observed a deformation of MIPS clusters from (standard) circular \cite{MIPS} to elliptic shapes. This effect is captured in Fig.\ \ref{fig:3}, where the width $w_\mathrm{c}$, height $h_\mathrm{c}$, and aspect ratio $w_\mathrm{c}/h_\mathrm{c}$ of the clusters are shown.
Interestingly, width and height change in such a way that the aspect ratio increases approximately linearly with $\nu$. This increase can be explained by the fact that, if $v_{\mathrm{A}}(\phi)$ is given by Eq.\ \eqref{eqn:v_A_n_fold} with $n=2$, the difference between the magnitudes of the active forces from the $y$-direction (proportional to $v_{\mathrm{A}}(\pi/2)$) and $x$-direction (proportional to $v_{\mathrm{A}}(0)$) grows linearly with $\nu$. 

\begin{figure*}[htpb]
\centering
\includegraphics[width=\linewidth]{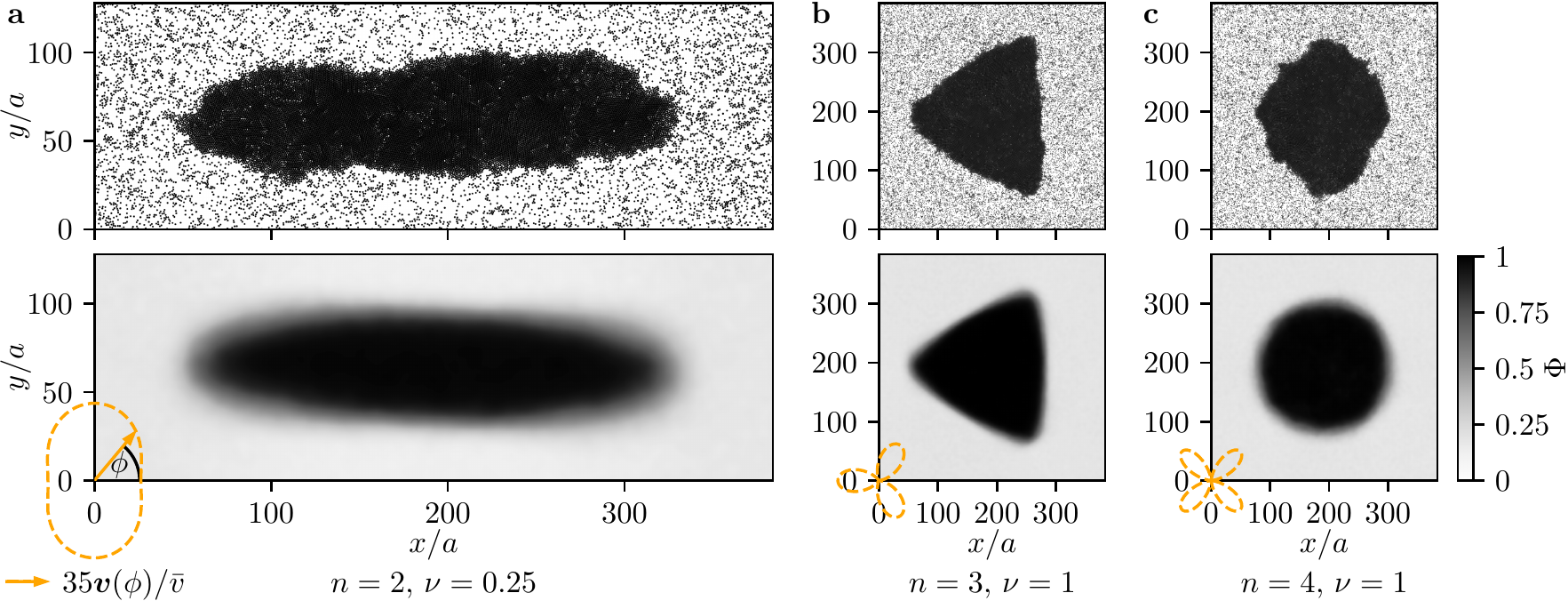}
\caption{\label{fig:4}Observed nonspherical cluster shapes that are self-assembled by ABPs with orientation-dependent propulsion speed $v_{\mathrm{A}, n}(\phi)$ (see Eq.\ \eqref{eqn:v_A_n_fold}) for different values of the symmetry order $n$ and angular modulation amplitude $\nu$. Snapshots of particle-based simulations (top) and simulation- and time-averaged local packing densities $\Phi(x,y)$ of the particles (bottom) are shown for (\textbf{a}) elliptic, (\textbf{b}) triangular, and (\textbf{c}) rectangular clusters.}
\end{figure*}

This leads to the question whether one can use orientation-dependent propulsion speeds to induce clusters of arbitrary intended shapes in ABP systems. To follow this idea, we investigated also the cases $n=3$ and $n=4$, choosing $\nu=0.25$ for $n=2$ to avoid unreasonably large aspect ratios of the clusters and $\nu=1$ otherwise to maximize the deformation phenomenon. 
The results are shown in Fig.\ \ref{fig:4}.
We observed elliptic clusters for $n=2$ (see Fig.\ \ref{fig:4}a), triangular clusters for $n=3$ (see Fig.\ \ref{fig:4}b), and rectangular clusters for $n=4$ (see Fig.\ \ref{fig:4}c). The elliptic and triangular clusters occur so reliably that they are also clearly visible when averaging the particle distribution in the stationary state over $11$ simulation runs and over time (bottom row of  Fig.\ \ref{fig:4}). In contrast, the rectangular clusters occur with their edges oriented either parallel to the coordinate axes or tilted by an angle $\pi/4$, such that this cluster shape is blurred by the averaging. Clusters with more complex shapes can be realized by a more complex dependence of $v_{\mathrm{A}}$ on $\phi$ (and possibly $\vec{r}$). 

These interesting findings may be utilized for the realization of programmable materials \cite{Stenhammar2016,Yan2016}. If the orientation-dependent propulsion can be controlled sufficiently well, the observed effect can make systems of active particles self-assemble into desired patterns. For example, electrically conducting ABPs may assemble switches where the elliptic clusters are used as bridges with certain orientations. The investigation of the programmable MIPS cluster shapes is our third main result.

\begin{figure}[htpb]
\centering
\includegraphics[width=\linewidth]{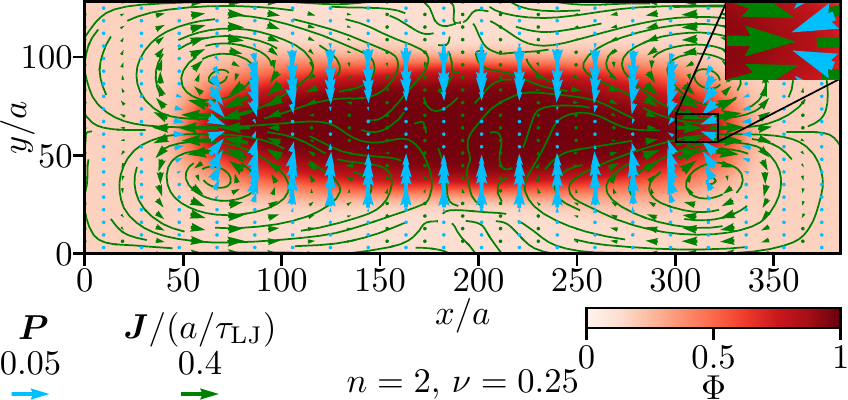}
\caption{\label{fig:6} Circulating currents in a stationary elliptic cluster for an orientation-dependent propulsion speed $v_{\mathrm{A}, n}(\phi)$ with $n=2$ and $\nu = 0.25$ (see Eq.\ \eqref{eqn:v_A_n_fold}).}
\end{figure}

Interestingly, the model \eqref{eqn:Continuity_equation} allows for circulating particle currents for $\underline{\mu}^{(2)}\neq\underline{0}$ since the curl of the current does not generally vanish in this case. Such currents are a central feature of active matter \cite{Cates2019}. For $v_{\mathrm{A}}=v_{\mathrm{A},2}$, Eq.\ \eqref{eqn:Continuity_equation} involves terms $\propto (\partial_{x}^2 - \partial_{y}^2) \rho$, which imply fluxes towards the denser region in the $y$-direction and out of it in the $x$-direction. This is related to the unusual MIPS cluster formation, since MIPS arises once particle fluxes point towards denser regions \cite{BialkeLS2013}. We therefore measured the current $\vec{J}(\vec{r})$ and the polarization $\vec{P}(\vec{r})$ (see Ref.\ \cite{SupMat} for definitions) in simulations for $n=2$ to check whether circulating currents do in fact exist in the elliptic steady state. The results, which confirm this expectation, are shown in Fig.\ \ref{fig:6}. Particles flow into the cluster from the $y$-direction (large $v_{\mathrm{A}}$) and out of it in the $x$-direction (small $v_{\mathrm{A}}$). This behavior is in stark contrast to the one known from standard ABPs, where there is no particle current through the interface in steady state \cite{Solon2015a,Tjhung2018}. Notably, $\vec{P}$ does always point inwards at the boundary of the ellipse, such that $\vec{J}$ and  $\vec{P}$ have opposite directions for $\phi\in \{0,\pi\}$ (see inset in Fig.\ \ref{fig:6}). The particles on the left- and right-hand side are therefore pushed out of the cluster by interaction forces even though their self-propulsion force (parallel to $\vec{P}$) points towards the cluster. A video of the cluster formation and the resulting steady state (including circulating currents) can be found in Ref.\ \cite{SI}. The circulating currents are our fourth main result.
 
In summary, we have shown that the collective behavior of ABPs with orientation-dependent propulsion gives rise to fascinating effects including tunable self-advection and circulating particle currents. An effective \Peclet{} number allows to map the spinodal for the anomalous MIPS observed here onto that of standard MIPS. The orientation-dependent propulsion can be employed to induce self-assembly of nonspherical clusters. Our findings provide new insights into the intriguing nonequilibrium dynamics of active particles and constitute an important step towards the realization of programmable materials by active soft matter. 

%\section*{Data availability}
%The data that supports the findings of this study are available within the article and its Supplementary Material \cite{SI}.

%\section*{Conflicts of interest}
%There are no conflicts of interest to declare.

M.t.V.\ thanks the Studienstiftung des deutschen Volkes for financial support. M.E.C.\ is funded by the Royal Society. R.W.\ is funded by the Deutsche Forschungsgemeinschaft (DFG, German Research Foundation) -- 283183152 (WI 4170/3-2). 
The simulations for this work were performed on the computer cluster PALMA II of the University of M\"unster. 

\noindent See Ref.\ \cite{SI} for the raw data for all figures.

\noindent S.B.\ and J.B.\ contributed equally to this work.

\bibliography{refs}
\end{document}